\input psfig.tex
\documentstyle[12pt,aasms4]{article}
\def\simlt{\lower.5ex\hbox{$\; \buildrel < \over \sim \;$}}
\def\simgt{\lower.5ex\hbox{$\; \buildrel > \over \sim \;$}}

\lefthead{Spaans \& Silk}
\righthead{The Equation of State of Interstellar Clouds}

\begin{document}

\title{The Polytropic Equation of State of Interstellar Gas Clouds}

\author{Marco Spaans\footnote{Hubble Fellow}}

\affil{Harvard-Smithsonian Center for Astrophysics, 60 Garden Street MS 51,
Cambridge, MA 02138}

\author{Joseph Silk}

\affil{Astrophysics, Department of Physics, University of Oxford,
Keble Road, Oxford OX1 3RH, U.K.\footnote{current address} and
Departments of Astronomy and Physics, University of California, 601
Campbell Hall, Berkeley, CA 94720}

\begin{abstract}
Models are presented for the polytropic equation of state of self-gravitating,
quiescent interstellar gas clouds.
A detailed analysis, including chemistry, thermal balance, and radiative
transfer, is performed
for the physical state of the gas as a function of density, metallicity,
velocity field, and background radiation field. It is found that the stiffness
of the equation of state strongly depends on all these physical parameters, and
the adiabatic index varies between $\sim 0.2-1.4$. The implications for star formation, in
particular at high redshift and in starburst galaxies, and the initial
stellar mass function are discussed.
\end{abstract}

{\it subject headings}: ISM: clouds - ISM: molecular process - radiative transfer - cosmology: theory - galaxies: starburst

\section{Introduction}

A fundamental problem in the theory of star formation is the physical structure
of molecular gas clouds, such as  dense cores and Bok globules, from which
(low-mass) stars are formed. Early work on the thermal and chemical balance
of interstellar clouds has been performed by de Jong, Dalgarno \& Boland (1980)
and Falgarone \& Puget (1985). A framework for active regions such as Orion
has been developed by Tielens \& Hollenbach (1985). The above work has been
extended and refined by many authors (see Hollenbach \& Tielens 1999 for a
review), driven to a large extent by recent advances in observational
techniques.

Examples of such observational progress include the catalog of 248 optically
selected Bok globules of Clemens \& Barvainis (1988), the NH$_3$ survey of
isolated cores in Taurus by Benson \& Myers (1989, and references therein),
the measurement of non-thermal line widths in regions surrounding ammonia
cores (Fuller \& Myers 1992, Goodman et al.\ 1998),
and the use of spectroscopy of various atoms and molecules as temperature and
density diagnostics of different types of interstellar clouds as summarized in
van Dishoeck (1997).

Despite this wealth of observational data, the nature of their equation of state (EOS) remains
a major theoretical problem that 
concerns the stability and collapse of these molecular clouds  The stiffness of the EOS can be
largely responsible for the resulting density probability function of
interstellar gas in the turbulent ISM (Scalo et al.\ (1998) and Passot \&
Vazquez-Semadeni 1998).
In particular, the value of gamma affects the densities that can be attained
behind shocks (Vazquez-Semadeni, Passot \& Pouquet 1996), the fraction of gas
driven to high densities by turbulent interactions (c.f.\ Scalo et al. 1998),
and the stability of clouds being overrun by shocks (e.g.\ Tohline, Bodenheimer
\& Christodoulou 1987; Foster \& Boss 1996).
In the present paper we concentrate on the effects of the EOS for the
fragmentation of individual collapsing clouds.
Before embarking on an analysis of the EOS of interstellar gas clouds, say of
the form $P=P(\rho ,T_{\rm g},T_{\rm d},V,B,I,C)$ for pressure $P$, density
$\rho$, gas temperatures $T_{\rm g}$, dust temperature $T_{\rm d}$, velocity
$V$, magnetic field $B$, radiation intensity $I$, and chemical composition $C$,
one should assess which of the possible physical quantities entering the EOS
dominates in  the parameter ranges under study.

The importance of $\rho$ and $T_{\rm g}$ is self-explanatory. The presence of
warm dust can play a role since it can heat the gas through gas-grain coupling
and radiatively through optically thick molecular lines (Takahashi, Hollenbach
\& Silk 1983). The velocity field, even when it does not contribute to the turbulent
pressure, strongly influences the optical depth of atomic and molecular cooling
lines, and hence the thermal balance of the medium. The magnetic field plays
an important role in the support of interstellar gas (McKee 1999). The presence
of ultraviolet or other hard radiation sources strongly influences the
chemical, thermal and ionization balance of the ambient medium, and hence the
overall EOS. Finally, the chemical composition of the gas also depends on the
metallicity since it is the abundance of atoms and molecules which influences
the local cooling rate.

In the present work, a polytropic equation of state, $P=K\rho^\gamma$, is
assumed to describe the physical state of the gas. The exponent $\gamma$ is
considered to be a function $\gamma (T_{\rm g},T_{\rm d},V,I,C)$. That is, the
magnetic pressure is assumed to be not dominant in the support of the model
cloud. The radiation field $I$, in addition to the radiation emitted by the
dust grains, is assumed to be given by the cosmic ray ionization rate.
In other words,
the visual extinction of the model cloud is assumed to be sufficient to shield
the gas from any ultraviolet (stellar) sources.

We note that $\gamma$ is the
logarithmic derivative of $P$ with respect to $\rho$. That is,
since $P\sim\rho T$ it follows that
$\gamma = 1+{{d {\rm log}T}\over{d{\rm log}\rho}}$. Through implicit
differentiation the above derivative is
related to the logarithmic derivatives of the heating and cooling
functions, which can be quite complex. It is this dependence of $\gamma$ on
the details of heating and cooling that is the main topic of the present paper.

In the remainder of the paper, a useful guide is the density dependence of
the heating and cooling rate. If the density dependence of the heating rate
is steeper (shallower) than that of the cooling rate, then $\gamma$ is
larger (smaller) than unity. A similar result holds for the kinetic
temperature. Here, the cosmic ray ionization rate is always linear in the
density and independent of temperature, whereas gas-grain heating and cooling
is quadratic in the density and proportional to
$T_{\rm g}^{1/2}(T_{\rm g}-T_{\rm d})$.
Optically thin, subthermally excited cooling
lines have a quadratic density dependence which becomes linear in the
thermalized (high density) limit and even effectively sub-linear when
radiative trapping
(optical depth effects) becomes important. Finally, collisional de-excitation
rates are characterized by powers of the temperature on the order of
$\sim 0.5-1.5$, whereas excitation rates are modified by an exponential
factor which is of the order of unity for kinetic temperatures larger than the
pertaining excitation temperatures.

In general, our results are applicable mostly to dense and/or well shielded
molecular clouds/cores where the pressure appears to be close to thermal, with
only a small additional contribution to the line-widths
from turbulent motions, and the ionization
fraction is small (Goodman et al.\ 1998, and references therein). That is,
we assume that the clouds are outside the regime where the so-called Larson
(1981) laws apply, which indicate an increase in the non-thermal line width
with length scale.

One should note that polytropic models can also be extended into the regime
where there is a significant contribution to the pressure from turbulence and
magnetic fields. The polytropic temperature $T$ should then be interpreted as
representing the total velocity dispersion $\sigma$, i.e., $T\propto\sigma^2$
(Maloney 1988). A complete description of polytropic models, including
composite ones for non-thermal pressure effects and bulk properties like
mass, radius and density contrast, is presented in Curry \&
McKee (1999). The aim here is to study the thermal regime and to investigate
the specific effects of metallicity, radiative transfer and internal sources
on the stiffness of the resulting local EOS.

\section{Model Description}

The results presented here were obtained by application of the numerical
code of Spaans (1996), described in detail in Spaans \& Norman (1997);
Spaans \& van Dishoeck (1997); and Spaans \& Carollo (1998). The interested
reader is referred to these papers for a description of the
underlying algorithm. The code has been specifically designed to solve large
chemical networks with a self-consistent treatment of the thermal balance for
all heating and cooling processes known to be of importance, and to be
computable with some sense of rigor, in the interstellar medium
(Spaans \& Ehrenfreund 1999).
Care has to be taken in the treatment of line trapping for the
atomic and molecular cooling lines which dominate the thermal balance.
Collisional de-excitation and radiative trapping is most
prominent in the density range $\sim10^3-10^4$ cm$^{-3}$ (Scalo et al.\ 1998). 
Our result were found to be in good agreement with those of
Neufeld et al.\ (1995)
for their, and our, adopted linear velocity gradient model (see further 
details below).

The adopted chemical network is based on the UMIST compillation (Millar,
Farquhar \& Willacy 1997) and is well suited
for low temperature dense molecular clouds, with the latest rates for the
important neutral-neutral reactions. The radiative transfer is solved by means
of a Monte Carlo approach (Spaans \& van Langevelde 1992)
with checks provided by an escape probability method
for large line optical depths. The thermal balance includes heating by cosmic
rays, absorption of infrared photons by H$_2$O molecules and their subsequent
collisional de-excitation, and gas-grain heating. The cooling includes 
the coupling between gas and dust grains (Hollenbach \& McKee 1989), atomic
lines of all metals, molecular lines, and all major isotopes, as described in
Spaans \& Norman (1997) and Spaans et al.\ (1994), and also in Neufeld,
Lepp \& Melnick (1995) for the regime of very large line optical depth.
All level populations are computed in statistical equilibrium at a relative
accuracy, also imposed on the ambient line radiation field, of no less
than $10^{-3}$. The combined chemical and thermal balance is required to obey a
convergence criterion between successive iterations of 0.5\%.

For the low metallicity computations, the relative contributions to the total
cooling by H$_2$ and HD can become significant. Since the value of $\gamma$
depends on the logarithmic derivative of the cooling rate with respect to
temperature and density, the end result could become quite sensitive to the
(quantum mechanical) treatment of H-H$_2$, H$_2$-H$_2$ and He-H$_2$ collional
processes.

The main differences between calculations of H-H2 and H2-H2 collisions are due
to different potentials (surfaces) and whether reactive collisions, i.e., the
exchange of an H atom, are included (important above 1000 K).
The shape of the potential surface is quite important for the lower
temperatures, where rotational quantum effects dominate. This situation has
improved greatly over recent years (c.f.\ Forrey et al.\ 1997).
Nevertheless, we have restricted our calculations to $Z\ge 0.01$ Z$_\odot$.
In this regime, there are other atomic and molecular coolants which still
contribute, at the 30\% level, diminishing the relative importance of the
aforementioned sensitivities.

As an additional check, we have performed runs with the recent le Bourlot et
al.\ (1999) H$_2$ rates. We have found no significant changes in $\gamma$ for
our adopted metallicities. Finally, for zero metallicity gas, and in the
absence of HD cooling, we found (albeit tentatively given the above caveats)
that $\gamma$ is always less than unity. The inclusion of HD restricts this
result to densities less than $\sim 10^5$ cm$^{-3}$.

\section{The Initial Mass Function} 

Before discussing the results of our computations, we want to outline how the
value of $\gamma$ could related to the process of star formation, i.e., how
a large value of $\gamma$ might suppress the formation of stars.

The polytropic equation
of state of interstellar gas clouds may be relevant to determinations of 
stellar masses. Our starting point is to pose the question: is there a
characteristic stellar mass? Observationally, the answer appears to be
positive. 
The IMF, $dN/dM\propto m^{-1-x}$, for the number $N$ of stars per unit of mass
$M$, has $x>1$ above $\sim 1 \rm M_\odot.$ Below $\sim 0.3 \rm M_\odot$,
virtually all data suggests a flattening to $-1\simlt x\simlt 0.$
The critical value that  defines a characteristic mass is $x \approx 1$
for a logarithmic divergence in total mass toward decreasing mass,
and there is no doubt that there is a transition from $x>1$ to $x<1$ in the
range $0.2-1\,\rm M_\odot$. Hence there is a characteristic stellar mass.
Whether this varies with environment is uncertain, but
$m_{\rm char}\approx 0.3\,\rm M_\odot$ fits most data (cf.\ Scalo 1998).

The status of theoretical discussions of a critical mass often centers
on the thermal Jeans mass, $m_{\rm J}\propto T^{3/2}\rho^{-1/2}$,
generalized to include external pressure, $m_{\rm J,P}\propto
T^2P^{-1/2}$, turbulence $m_{\rm J,t}\propto \Delta V^{4}P^{-1/2}$, or
magnetic flux support $m_{\rm J,B}\propto B^3\rho^{-1}$
(c.f.\ Elmegreen 1999, Larson 1998).
The generalized Jeans scale most likely applies to the masses of atomic and
molecular clouds or to self-gravitating clumps. However, the fact that in the
interstellar medium, the Jeans mass in diffuse atomic or molecular
clouds generally exceeds stellar masses by a large factor, apart from
the very coldest clumps, argues that the physics of the IMF is more
complicated than Jeans instability.  The Jeans mass is a fragmentation
scale according to linear theory, and three-dimensional simulations
indeed find that clump masses are described by the Jeans
scale. Nevertheless, simulations cannot currently resolve the scales of
individual protostars on solar mass scales.
Analytic arguments, on the contrary, of opacity-limited
fragmentation find scales of $\sim 0.01\,\rm M_\odot$, or far less
than the characteristic stellar mass scale.

All of this seems to suggest that the Jeans mass is not very
relevant to the IMF, not
withstanding the Elmegreen (1999) and Larson (1998) results.
If it were, of course, the critical polytropic index
for fragmentation to occur is 4/3.
This has the following significance: one can write $m_{\rm J}\propto
\rho^{(3/2)(\gamma-4/3)}$.
As collapse occurs, smaller and smaller masses fragment, and one plausibly
ends up with a powerlaw IMF only if $\gamma <4/3$.
This argument could apply to the clump mass function within molecular clouds.
The physics of nonlinear fragmentation is complicated, and this is at least a
necessary condition for a power law IMF to continue to very small mass scales
(compared to the initial Jeans mass).

But what really determines stellar masses? Physical processes that
must play a role in determining the characteristic stellar mass scale
include clump coagulation (Allen \& Bastien 1995),
clump interactions (Klessen, Burkert \& Bate
1998) and protostellar outflows (Nakano, Hasegawa \& Norman 1995; Silk 1995)
that halt accretion. One conjecture is that feedback
from protostars plays a crucial role in limiting accretion.
The latter depends crucially on the
accretion rate to protostellar cores (various authors relate inflow to
outflow rates on energetic grounds), which in turn depends sensitively
on the turbulent
velocity. The maximum embedded young stellar object luminosity correlates
with NH$_3$ core line width $\Delta V$ (Myers \& Fuller 1993) and clump mass
(Kawamura et al.\ 1998) over a wide range. Myers \& Fuller (1993) concluded
that variations in the initial velocity dispersion from core to core is a key
determinant of the stellar mass formed from such a core. Also, Fuller \& Myers
(1992) concluded that the physical basis of line width-size relations (the
so-called Larson laws) is part of
the initial conditions for star formation. The theoretical
justification incorporates disk physics and, at least conceptually, magnetic
fields (Adams \& Fatuzzo 1996; Silk 1995).
So if the above holds (theoretically and observationally,
both aspects are relevant), then one might argue that the characteristic mass
of a young stellar object depends primarily on $\Delta V$. The latter is
determined by the kinetic temperature if the thermal line width dominates in a
prestellar core. This situation holds for low mass star formation but less so
for the high mass cores, where the nonthermal contribution to the line width
(i.e.\ turbulent or magnetic) can be as large as 50-80\% (c.f.\ Myers \& Fuller
1993). In any case, $\gamma =1$ appears the most natural critical value of
interest.

If $\gamma =1$ is indeed a critical value in this sense, there is the
possibility that the primordial EOS with a stiffer polytropic index,
$\gamma >1$, leads to a peaked IMF, i.e.\ one biased toward massive stars,
whereas current star formation, where the molecular gas is characterised by an
EOS with $\gamma <1$, results in runaway fragmentation and hence a power law
IMF.

\section{Results}

The model results\footnote{Extensive tables can be obtained from MS.} are
obtained for self-gravitating, quiescent spherical
clouds with total column densities per unit of velocity
given by
$$N^{\rm SIS}({\rm H}_2)=5.1\times 10^{19} n({\rm H}_2)^{1/2}\quad
{\rm cm}^{-2}\ {\rm per}\ {\rm km}\ {\rm s}^{-1},\eqno(1)$$
(the singular isothermal sphere value; Neufeld et al.\ 1995) to
set the optical depth parameter for any species, total
hydrogen densities $n_{\rm H}=10^2-10^6$ cm$^{-3}$, metallicities
$Z=1-0.01Z_\odot$ in solar units,
and a possible homogeneous background infrared radiation (IR) field given by
100 K dust grains. The total visual extinction through the region containing
the model cloud is fixed at $A_{\rm V}=100$ mag, although this is only
important for  determining the dust emission optical depth in the IR
background case. Models are also constructed for a constant velocity gradient
field of 3 km s$^{-1}$ pc$^{-1}$, which is a factor of three larger than the
value adopted in Goldsmith \& Langer (1978), corresponding to an optical depth
parameter of
$$N^{\rm VG}({\rm H}_2)=1.0\times 10^{18} n({\rm H}_2)\quad {\rm cm}^{-2}
\ {\rm per}\ {\rm km}\ {\rm s}^{-1}.\eqno(2)$$
It has been shown by Neufeld \& Kaufman (1993) that this choice violates the
virial theorem in the limit of high density. Therefore, (1) is interpreted 
to apply to a (statistically) static cloud supported by thermal pressure,
whereas (2) should be seen as representative of a dynamic medium in a state of
infall or outflow. The cosmic ray ionization rate is taken equal to
$\zeta_{\rm CR}=3\times 10^{-17}$ s$^{-1}$.

\subsection{Variations in $\gamma$: Competition between Heating and Cooling Processes}

Figures 1 and 2 present results which do not include an IR background
radiation field, e.g.\ little or no ongoing star formation, or a warm CMB.
Figure 1 displays the results for a quiescent velocity field with a
velocity dispersion of $\Delta V=0.3$ km s$^{-1}$ for H$_2$. Figure 2
is for the
corresponding case of a velocity gradient of 3 km s$^{-1}$ pc$^{-1}$.
The value of $\gamma$ becomes of the order of unity for low metallicities,
$Z<0.03Z_\odot$, and densities $n_{\rm H}<10^5$ cm$^{-3}$. The reason is that
the cooling of the ambient medium becomes dominated by H$_2$ and HD
(Norman \& Spaans 1997), which emit optically thin line radiation.
At low densities, the level populations are not thermalized and the cooling
rate has a density dependence which is steeper than linear and typically
quadratic, whereas the cosmic ray heating rate is linear in the density.

At higher densities gas-grain cooling becomes
important. This has a quadratic density dependence, and the equation
of state softens in the absence of any embedded
sources (Scalo et al.\ 1998), but to a lesser extent due to the lower
dust abundance, for small metallicities. Even though no computations were
performed beyond densities of $\sim 3\times 10^6$ cm$^{-3}$, the gas-grain
coupling at very large densities is so strong that the gas temperature follows
the grain temperature and the value of $\gamma$ will increase again and attain
a value of unity if embedded protostars are present and the ambient radiation
field is not coupled to the local gas density (see below).

The high $\gamma$ region between $\sim 10^3$ and $\sim 10^4$ cm$^{-3}$ is the
result of line trapping and collisional de-excitation. Line trapping causes
the cooling rate to depend on the density with an effective power less than
unity due to absorption and subsequent collisional de-excitation,
while the cosmic ray heating rate is always linear in the density, independent
of the kinetic temperature.
It should be emphasized that the effect of radiative trapping is stronger for
the quiescent velocity field, but also depends crucially on the specific
implementation that we have chosen. A truly turbulent
velocity field (Kegel, Piehler \& Albrecht 1993), i.e.\ one with a finite
correlation length, may lead to a quite different value of $\gamma$ in the
density range $\sim 10^3-10^4$ cm$^{-3}$.

\subsection{High Redshift Star formation}

As can be seen in Figure 3, the inclusion of an IR background leads to a much
stiffer polytropic equation of state for high, $Z>0.1Z_\odot$, metallicities.
The reason is that the abundant presence of water leads to heating through
absorption of IR radiation and subsequent collisional de-excitation
(Takahashi et al.\ 1983) for densities between $\sim 3\times 10^3$ and
$\sim 3\times 10^4$
cm$^{-3}$. Temperature effects play a role here in the
determination of the line opacities, but this is essentially a heating term
that is between linear and quadratic in the ambient density.
Since this heating rate is roughly proportional to
the abundance of H$_2$O, it is negligible for metallicities $Z<0.1Z_\odot$.
At very high densities, $n_{\rm H}>3\times 10^6$ cm$^{-3}$, the gas and dust
become completely coupled, and $\gamma$ approaches unity since the gas
temperature
will simply follow the dust temperature. This only follows provided that the
strength of the IR radiation field is independent of the ambient density.
This point will re-addressed in the next subsection on starburst galaxies.

Note also that $\gamma$ decreases to unity from above for densities larger than
$\sim 10^5$ cm$^{-3}$ because of the high temperature of the
FIR background. This causes
the dust to heat the gas and results in a relatively stiff EOS, even for low
metallicities where H$_2$ and HD dominate. This is because at these high
densities the overall cooling rates are at their thermodynamic limit, roughly
linear in the density, while the dust-gas heating rate goes like the square of
the density.
Note that for $T<200$ K, HD quickly becomes the dominant
coolant of pristine gas. The lower equilibrium temperature of the gas
facilitates the thermalization of the level populations. Finally, for
densities larger than $3\times 10^5$ cm$^{-3}$ radiative trapping becomes
important for HD, leading to a density dependence of its cooling rate of
less than unity.

Even though no results are shown for intermediate (30-60 K) temperature dust
backgrounds, $\gamma$ first
decreases to somewhat below unity around $10^5$ cm$^{-3}$ in these cases, if
the metallicity is less than 0.1 Z$_\odot$. It
then approaches unity from below for the reasons given above.
Finally, in the limit of a large velocity gradient, H$_2$O plays a minor role
and its heating (or cooling) does not strongly influence the thermal balance
of the gas (Takahashi et al.\ 1983).

Since an IR background is naturally present in the form of the
CMB for redshifts larger than $z\sim 10-30$, one
can speculate on the nature of very early star formation.
That is, any dense shielded region which has been enriched by metals through
the first supernova explosions, viewed here as the product of the very first
population III stars, will become more stable in the presence of an intense CMB
background. Conversely,
metal-poor regions will retain a polytropic exponent of $\approx 1$. Therefore,
high redshift star formation may exhibit the counter-intuitive property that
metal enrichment {\it and} a warm CMB together ``halt'' the process of star
formation for a metallicity $Z_{\rm c}> 0.1Z_\odot$, even though the gas is
intrinsically capable of cooling efficiently (Spaans \& Norman 1997).

Once the CMB temperature drops below $\sim 40$ K, at $z\sim 15$, any dense,
$n_{\rm H}>10^4$ cm$^{-3}$, cloud of high or even modest metallicity rapidly develops a
$\gamma$-value smaller than unity and is prone to collapse. In effect, although more
detailed calculations are required to substantiate this, the CMB may play an
integral part in determing the epoch of efficient early star formation. One
also sees that the primordial IMF can change strongly as one goes from a
$\gamma >1$ to a $\gamma <1$ EOS. In the former regime one expects only large
density excursions to collapse, leading to an IMF biased toward massive stars.

\subsection{Starburst Galaxies}

A similar phenomenon can occur in luminous (IR) starburst galaxies in
that warm, $\sim 50-200$ K, opaque dust hardens the polytropic equation of
state of a nuclear starburst region. The system Arp 220 is a possible example,
being optically thick at 100 $\mu$m. From the results presented here, one
concludes that such IR luminous opaque systems have an IMF which is
quite shallow, favoring high mass star formation. This may in fact
be desirable
to maintain the required power of such galaxies through supernova explosions
(c.f.\ Doane \& Mathews 1993).

Note though that in these extreme systems there is a strong coupling between
the ambient radiation field and the star formation rate $R$, and hence with
the gas density. Typical Schmidt star formation laws adopt an observationally
motivated value of $1-2$ for the dependence of $R$ on gas density (Kennicutt
1998) in star-forming spiral galaxies, with a best fit value of $r=1.4$. Since
the resulting dust temperature depends on the $1/5$ power of the ultraviolet
energy density (Tielens \& Hollenbach 1985), the effective $\gamma$ for $r=2$
($r=1$) is 1.4 (1.2), consistent with the above results.

\section{Discussion}

The polytropic equation of state of interstellar gas clouds has been computed
for a range of physical parameters, and its relation to the IMF, high redshift
star formation, and starburst galaxies has been investigated. A much needed
extension of the results presented here lies in MHD, and particularly
turbulence. The magnetic and velocity fields are important for pressure
support away from the most centrally condensed part of a molecular cloud core,
and radiative transfer effects in a turbulent medium are non-trivial.
More to the point, the polytropic EOS indices computed here, with their
relevance to the IMF as discussed in Section 4, are to be applied to a medium
with a given cloud density probability distribution, as investigated
in Scalo et al.\ (1998). Such a cloud density probability distribution can
only be derived from detailed hydrodynamic simulations which include magnetic
pressure and Coriolis forces.
Furthermore, $\gamma$ can also affect, in a way still to be investigated,
star formation and the underlying IMF in other scenarios like
turbulence-induced condensation or shocked-cloud induced star formation.
We would like to mention as well the work by Ogino, Tomisaka \& Nakamura (1999)
that shows that the mass accretion rate in their polytropic simulations of
collapsing clouds depends a lot on whether gamma is taken as 0.8 or 1.2.

Extensions in terms of multiple polytropic models for the core and envelope of
a molecular cloud, such as discussed in Curry \& McKee (1999), are another
avenue to explore for the bulk properties of interstellar clouds. Caveats in
the approach presented here are grain mantle evaporation which can influence
the abundance of H$_2$O and hence the heating rate in the presence of an IR
background, uncertainties in the chemical reaction rates of neutral-neutral
reactions at low temperatures, and details of the freeze-out onto grains of
molecular species. We hope that through a synthesis of the approach adopted
here and those of others mentioned above, one can address the question of cloud
stability and collapse in a more general (cosmological) framework.

\acknowledgments
We are grateful to the referee, John Scalo, for his constructive comments,
which improved the presentation of this paper, and his emphasis on the
computation of the zero metallicity case.
Discussions with Ewine van Dishoeck and John Black on the collisional
cross sections of H$_2$ are also gratefully acknowledged.
MS is supported by NASA through grant HF-01101.01-97A,
awarded by the Space Telescope Institute, which is
operated by the Association of Universities for Research in Astronomy,
Inc., for NASA under contract NAS 5-26555. J.S. has been  supported in part by grants
from NSF and NASA at UC Berkeley.

\newpage

\begin{figure}
\label{figure1}
\caption{Density and metallicity dependence of the polytropic index for
model I, discussed in the text, for no IR background and a
quiescent velocity field with $\Delta V=0.3$ km s$^{-1}$.}
\end{figure}

\begin{figure}
\label{figure2}
\caption{Density and metallicity dependence of the polytropic index for
model II, discussed in the text, for no IR background and a spherically
symmetric constant velocity gradient field of 3 km s$^{-1}$ pc$^{-1}$.}
\end{figure}

\begin{figure}
\label{figure3}
\caption{Density and metallicity dependence of the polytropic index for
model I, discussed in the text, for a $T_{\rm d}=100$ K IR background
from dust, $A_{\rm V}=100$ mag, and a quiescent velocity field.}
\end{figure}

\clearpage
\centerline{\psfig{figure=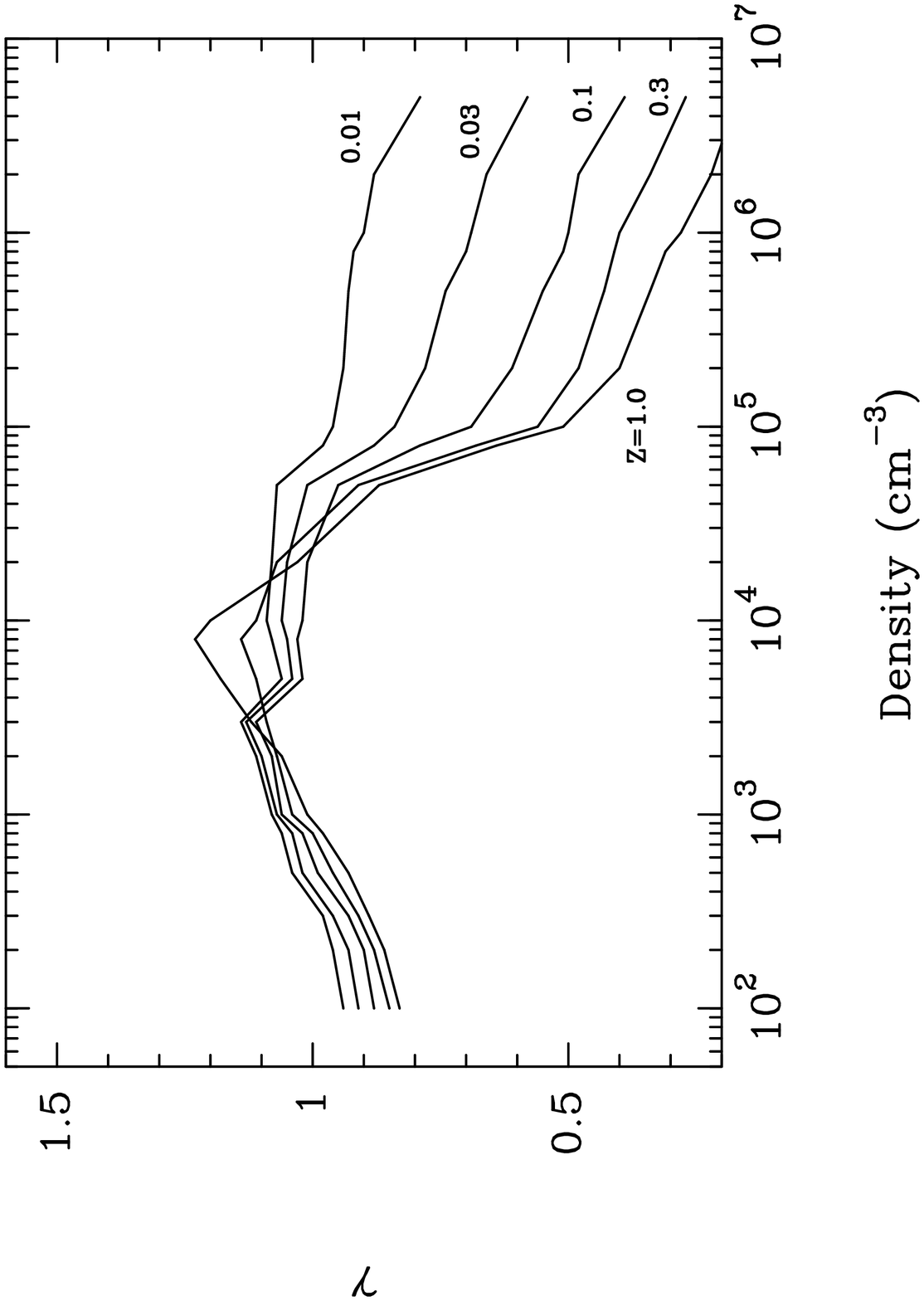,width=15.0truecm}}

\clearpage
\centerline{\psfig{figure=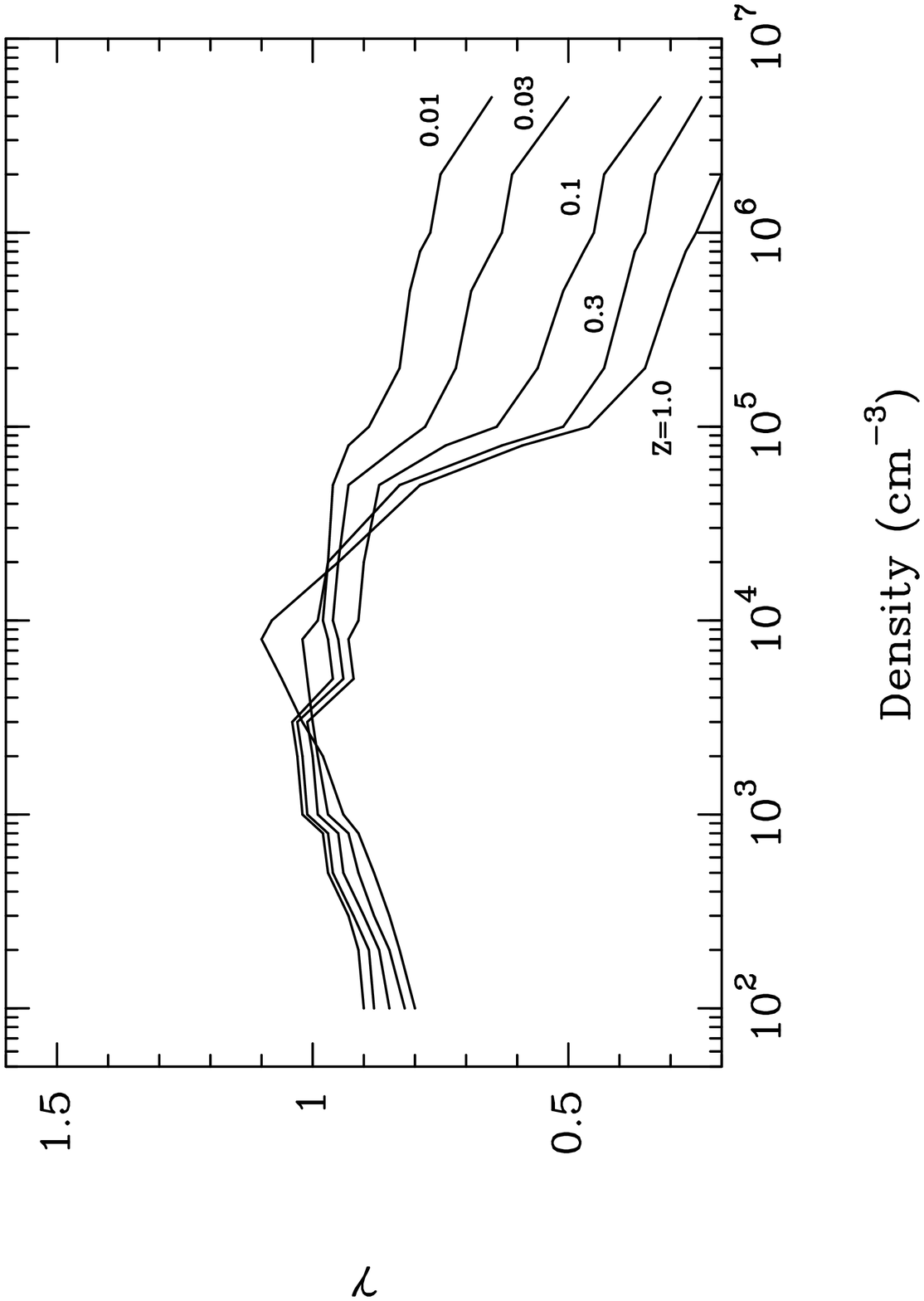,width=15.0truecm}}

\clearpage
\centerline{\psfig{figure=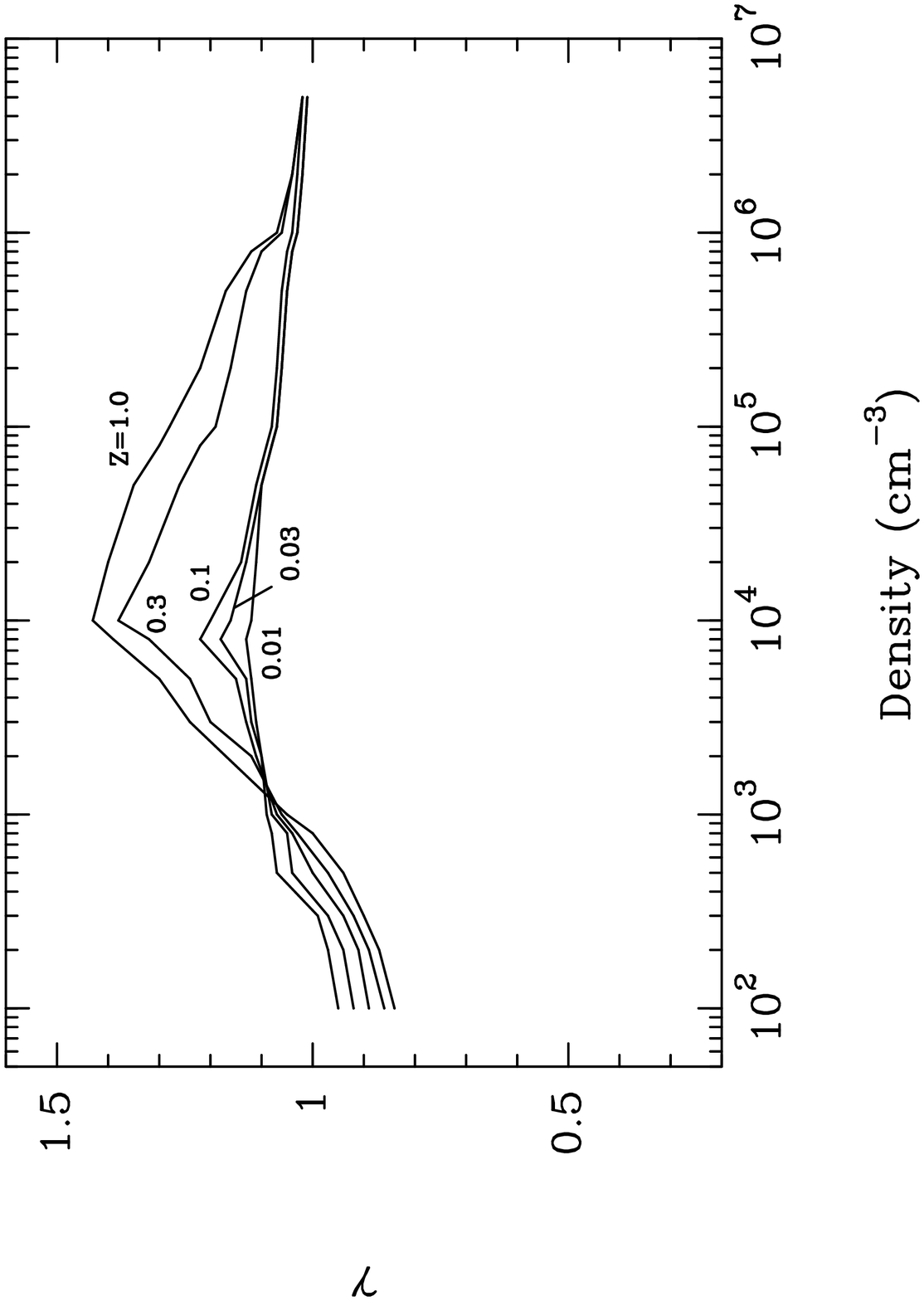,width=15.0truecm}}


\newpage
\begin{thebibliography}{}
\bibitem[]{}Adams, F.C. \& Fatuzzo, M.\ 1996, ApJ, 464, 256
\bibitem[]{}Allen, E.J. \& Bastien, P.\ 1995, ApJ, 452, 652
\bibitem[]{}Benson, P.J. \& Myers, P.C.\ 1989, ApJS, 71, 89
\bibitem[]{}Clemens, D.P. \& Barvainis, R.\ 1988, ApJS, 68, 257
\bibitem[]{}Curry, C.L. \& McKee, C.F.\ 1999, astro-ph/9908071
\bibitem[]{}de Jong, T., Dalgarno, A. \& Boland W.\ 1980, A\&A, 91, 68
\bibitem[]{}Doane, J. \& Mathews, W.G.\ 1993, ApJ, 419, 573
\bibitem[]{}Elmegreen, B.G.\ 1999, ApJ, 515, 323
\bibitem[]{}Falgarone, E. \& Puget, J.-L.\ 1985, A\&A, 142, 157
\bibitem[]{}Forrey, R.C., Balakrishnan, N., Dalgarno, A. \& Lepp, S.\ 1997,
ApJ, 489, 1000
\bibitem[]{}Foster, P.N. \& Boss, A.P.\ 1996, ApJ, 468, 784
\bibitem[]{}Fuller, G.A. \& Myers, P.C.\ 1992, ApJ, 384, 523
\bibitem[]{}Goldsmith, P.F. \& Langer, W.D.\ 1978, ApJ, 222, 881
\bibitem[]{}Goodman, A.A., Barranco, J.A., Wilner, D.J. \& Heyer, M.H.\ 1998,
ApJ, 504, 223
\bibitem[]{}Hollenbach, D.A. \& McKee, C.F.\ 1989, ApJ, 342, 306
\bibitem[]{}Hollenbach, D.A. \& Tielens, A.G.G.M.\ 1999, Rev.\ Modern Phys.,
71, 173
\bibitem[]{}Kawamura, A., Onishi, T., Yonekura, Y., Dobashi, K., Mizuno, A.,
Agawa, H. \& Fukui, Y.\ 1998, ApJS, 117, 387
\bibitem[]{}Kegel, W.H., Piehler, G. \& Albrecht, M.A.\ 1993, A\&A, 270, 407
\bibitem[]{}Kennicutt, R.J., 1998, ApJ, 498, 541
\bibitem[]{}Klessen, R.S., Burkert, A. \& Bate, M.R.\ 1998, ApJ, 501, L205
\bibitem[]{}Larson, R.B.\ 1998, MNRAS, 301, 569
\bibitem[]{}Larson, R.B.\ 1981, MNRAS, 194, 809
\bibitem[]{}le Bourlot, J., Pineau des For\^ets, G. \& Flower, D.R., 1999,
MNRAS, 305, 802
\bibitem[]{}Maloney, P.\ 1988, ApJ, 334, 761
\bibitem[]{}McKee, C.F., 1999, astro-ph/9901370
\bibitem[]{}Millar, T.J., Farquhar, P.R.A. \& Willacy, K., 1997, A\&AS, 121,
139
\bibitem[]{}Myers, P.C. \& Fuller G.A.\ 1993, ApJ, 402, 635
\bibitem[]{}Nakano, T., Hasegawa, T. \& Norman, C.A.\ 1999, ApJ, 450, 183
\bibitem[]{}Neufeld, D.A. \& Kaufman, M.J.\ 1993, ApJ, 418, 263
\bibitem[]{}Neufeld, D.A., Lepp, S. \& Melnick, G.J.\ 1995, ApJS, 100, 132
\bibitem[]{}Norman, C.A. \& Spaans, M.\ 1997, ApJ, 480, 145
\bibitem[]{}Ogino, S., Tomisaka, K. \& Nakamura, F.\ 1999, PASJ, 51, 637
\bibitem[]{}Passot, T.\& Vazquez-Semadeni, E.\ 1999, Phys.\ Rev.\ E, 58, 4501
\bibitem[]{}Scalo, J. \ 1998, 
in {\it The Stellar Initial Mass Function} (38th Herstmonceux Conference)
ASP Conference Series, 
Vol. 142, 1998, ed. G. Gilmore and D. Howell.
\bibitem[]{}Scalo, J., V\'azquez-Semadeni, E., Chappell, D. \& Passot, T.\
1998, ApJ, 504, 835
\bibitem[]{}Silk, J.\ 1995, ApJ, 438, L41
\bibitem[]{}Spaans, M.\ 1996, A\&A, 307, 271
\bibitem[]{}Spaans, M. \& Ehrenfreund, P.\ 1999, in ``Laboratory Physics and
Space Research'', ASSL Vol.\ 236, eds.\ P.\ Ehrenfreund, K.\ Krafft, H.\
Kochan, and V.\ Pironello (Kluwer:Dordrecht), p.\ 1
\bibitem[]{}Spaans, M. \& Carollo, C.M.\ 1998, ApJ, 502, 640
\bibitem[]{}Spaans, M. \& van Dishoeck, E.F.\ 1997, A\&A, 323, 953
\bibitem[]{}Spaans, M. \& Norman 1997, ApJ, 488, 27
\bibitem[]{}Spaans, M., Tielens, A.G.G.M., van Dishoeck, E.F. \& Bakes, E.L.O.,
1994, ApJ, 437, 270
\bibitem[]{}Spaans, M. \& van Langevelde, H.J.\ 1992, MNRAS, 258, 159
\bibitem[]{}Takahashi, T., Hollenbach, D.J. \& Silk, J.\ 1983, ApJ, 275, 145
\bibitem[]{}Tielens, A.G.G.M. \& Hollenbach, D.A., 1985, ApJ, 291, 722
\bibitem[]{}Tohline, J.E., Bodenheimer, P.H. \& Christodoulou D.M.\ 1987, ApJ,
322,787
\bibitem[]{}van Dishoeck, E.F., ed.\ ``Molecules in Astrophysics: Probes and
Processes'', 1997, IAU Symp.\ 178, Kluwer
\bibitem[]{}Vazquez-Semadeni, E., Passot, T.\& Pouquet, A.\ 1996, ApJ, 473, 881
\end{thebibliography}
\end{document}